\documentstyle[amssymb,epsf,prd,aps,floats]{revtex}

\begin{document}
\title{Variations of Alpha in Space and Time }
\author{John D. Barrow$^1$, Jo\~ao Magueijo$^2$ \& H\aa vard Bunes Sandvik$^2$}
\address{$^1$DAMTP, Centre for Mathematical Sciences, \\
Cambridge University, Wilberforce Road,\\
Cambridge CB3 0WA, UK\\
$^2$Theoretical Physics, Blackett Laboratory, \\
Imperial College, Prince Consort Road, \\
London SW7 2BZ, UK}
\maketitle

\begin{abstract}
We study inhomogeneous cosmological variations in the fine structure
'constant', $\alpha ,$ in Friedmann universes. Inhomogeneous motions of the
scalar field driving changes in $\alpha $ display spatial oscillations that
decrease in amplitude with increasing time. The inhomogeneous evolution
quickly approaches that found for exact Friedmann universes. We prove a
theorem to show that oscillations of $\alpha $ in time (or redshift) cannot
occur in Friedmann universes in the BSBM theories considered here.
\end{abstract}

\pacs{PACS Numbers: *** }



\section{Introduction}

Elsewhere, \cite{sbm,bsm}, we have discussed the behaviour of a class of
cosmologies in an exact theory in which the fine structure ``constant''
varies in time. This theory of Sandvik, Barrow and Magueijo is an extension,
to include the self-gravitation of the dielectric medium, of Bekenstein's
prescription \cite{bek2} for generalising Maxwell's equations to incorporate
varying electron charge. We will refer to it as the BSBM theory. The fine
structure ``constant'' $\alpha $ varies through the space-time dynamics of a
scalar ```dielectric'' field $\psi ,$ (where $\ \alpha =\exp [2\psi ]$) in
these theories. However the overall behaviour is significantly affected by
the form of the coupling. Even though the requirement that the energy in $%
\psi $ be positive definite fixes the sign of the coupling constant $\omega $
we find that $\psi $ is driven by a term of the form ${\cal L}_{em}/\omega $%
, where ${\cal L}_{em}$ is the electromagnetic Lagrangian. In general, $%
{\cal L}_{em}$ can be positive or negative, a fact we parameterize in terms
of $\zeta ={\cal L}\ _{em}/\rho $, where $\rho $ is the energy density. The
sign of $\zeta $ for the dark matter in the universe turns out to be of
exceptional significance.

In our earlier studies \cite{sbm,bsm} we have focussed on the case where $%
\zeta <0$ and the dark matter in the universe is dominated by magnetic field
couplings. This was motivated by the discovery that $\zeta <0$ matter leads
to a slow (logarithmic){\it \ increase} in the value of $\alpha $ with
cosmic time during the matter era of the universe and constant behaviour
during any period in which the expansion is curvature dominated,
accelerates, or is dominated by radiation in universes with a
matter-radiation balance like our own. Thus for $\zeta <0$ we find slow
time-evolution of the fine structure 'constant' that is consistent with the
observations of Webb et al \cite{webb,webb2,murphy} of $\Delta \alpha
/\alpha =(-0.72\pm 0.18)\times 10^{-5}$ at $z=1-3.5.$ They are also
consistent with the low-redshift, $z\sim 0.1$, upper limits on time
variation of $\alpha $ provided by Oklo \cite{sh,fujii},and high-redshift
constraints imposed by the microwave background temperature fluctuations {\ 
\cite{avelino,bat},} and primordial nucleosynthesis \cite{bbn}. Other hints
of varying constants in astronomical studies have recently been reported by
Ivanchik et al \cite{ivan}.

All of the studies described above have been performed in the context of an
exact isotropic and homogeneous Friedmann universe. All variations in the
fine structure ``constant'' therefore depend only on cosmic time. However,
the rate of variation that is suggested by recent astronomical observations
of quasar spectra, or allowed by geophysical data at recent times, is very
small, $\Delta \alpha /\alpha \sim 10^{-5},$ and spatial variations in the
rate of time variation could easily be of similar order \cite{bt}. It is
therefore important to determine if spatial variations in the rate of change
of $\alpha $ are significant in the BSBM theory and whether they allow
different modes of time variation to occur in addition to those studied in
the purely homogeneous variations found in refs. \cite{bsm,sbm}.

\section{The BSBM theory}

There are a number of possible theories allowing for the variation of the
fine structure constant, $\alpha $. In the simplest cases one takes $c$ and $%
\ \hbar $ to be constants (see however \cite{covvsl,moffatal}) and
attributes variations in $\alpha $ to changes in the electron charge, $e,$
or the permittivity of free space (see ref. \cite{am} for a discussion of
the meaning of this choice). This is done by letting $e$ take on the value
of a real scalar field which varies in space and time (for more complicated
cases, resorting to complex fields undergoing spontaneous symmetry breaking,
see the case of fast tracks discussed in \cite{covvsl}). Thus $%
e_0\rightarrow e=e_0\epsilon (x^\mu ),$ where $\epsilon $ is a dimensionless
scalar field and $e_0$ is a constant denoting the present value of $e$. This
operation implies that some well established assumptions, like charge
conservation, must give way \cite{land}. Nevertheless, the principles of
local gauge invariance and causality are maintained, as is the scale
invariance of the $\epsilon $ field (under a suitable choice of dynamics).
In addition there is no conflict with local Lorentz invariance or
covariance. With this set up in mind, the dynamics of our theory is then
constructed as follows. Since $e$ is the electromagnetic coupling, the $%
\epsilon $ field couples to the gauge field as $\epsilon A_\mu $ in the
Lagrangian and the gauge transformation which leaves the action invariant is 
$\epsilon A_\mu \ \rightarrow \epsilon A_\mu +\chi _{,\mu },$ rather than
the usual $A_\mu \ \rightarrow A_\mu +\chi _{,\mu }.$ The gauge-invariant
electromagnetic field tensor is therefore 
\begin{equation}
F_{\mu \nu }=\frac 1\epsilon \left( (\epsilon A_\nu )_{,\mu }-(\epsilon
A_\mu )_{,\nu }\right) ,
\end{equation}
which reduces to the usual form when $\epsilon $ is constant. The
electromagnetic part of the action is still 
\begin{equation}
S_{em}=-{\frac 14}\int d^4x\sqrt{-g}F^{\mu \nu }F_{\mu \nu },
\end{equation}
and the dynamics of the $\epsilon $ field are controlled by the kinetic term 
\begin{equation}
S_\epsilon =-\frac 12\frac{\hbar c}{l^2}\int d^4x\sqrt{-g}\frac{\epsilon
_{,\mu }\epsilon ^{,\mu }}{\epsilon ^2},
\end{equation}
as in dilaton theories. Here, $l$ is the characteristic length scale of the
theory, introduced for dimensional reasons. This constant length scale gives
the scale down to which the electric field around a point charge is
accurately of Coulomb type. The corresponding energy scale, $\omega =\hbar
c/l,$ has to lie between a few tens of $MeV$ and Planck scale, $\sim
10^{19}GeV,$ to avoid conflict with experiment.

Our generalisation of the scalar theory proposed by Bekenstein \cite{bek2}
described in ref. \cite{sbm} includes the gravitational effects of $\ \psi
=\log \epsilon $. It gives the field equations: 
\begin{equation}
G_{\mu \nu }=8\pi G\left( T_{\mu \nu }^m+T_{\mu \nu }^\psi +T_{\mu \nu
}^{em}e^{-2\psi }\right) ,  \label{ein}
\end{equation}
where the various $T_{\mu \nu }$ are the matter, $\psi $ and electromagnetic
stress energy tensors. Recall the $\psi $ lagrangian is ${\cal L}_\psi =-{\ 
\frac \omega 2}\partial _\mu \psi \partial ^\mu \psi $ and the $\psi $ field
obeys the equation of motion 
\begin{equation}
\Box \psi =\frac 2\omega e^{-2\psi }{\cal L}_{em},  \label{boxpsi}
\end{equation}
where we have defined the coupling constant $\omega =(\hbar c)/l^2$. This
constant is of order $\sim 1$ if, as in \cite{sbm}, the energy scale is
similar to Planck scale. It is clear that ${\cal L}_{em}$ vanishes for a sea
of pure radiation since then ${{\cal L}_{em}}=(E^2-B^2)/2=0$. We therefore
expect the variation in $\alpha $ to be driven by electrostatic and
magnetostatic energy-components rather than electromagnetic radiation. In
order to make quantitative predictions we need to know how much of the
non-relativistic matter contributes to the RHS of Eqn.~(\ref{boxpsi}). This
is parametrised by $\zeta \equiv {\cal L}_{em}/\rho $, where $\rho $ is the
energy density, and for baryonic matter ${{\cal L}_{em}}=E^2/2$. For protons
and neutrons $\zeta _p$ and $\zeta _n$ can be {\it estimated} from the
electromagnetic corrections to the nucleon mass, $0.63$ MeV and $\ -0.13$
MeV, respectively \cite{zal,olive}, . This correction contains the $\ E^2/2$
contribution (always positive), but also terms of the form $j_\mu \ a^\mu $
(where $j_\mu $ is the quarks' current) and so cannot be used directly.
Hence we take a guiding value $\zeta _p\approx \zeta _n\sim \ 10^{-4}$.
Furthermore the cosmological value of $\zeta $ (denoted $\zeta _m$) has to
be weighted by the fraction of matter that is non-baryonic, a point ignored
in the literature \cite{bek2,livio}. Hence, $\zeta _m$ depends strongly on
the nature of the dark matter and can take both positive and negative values
depending on which of Coulomb-energy or magnetostatic energy dominates the
dark matter of the Universe. It could be that $\zeta \ _{CDM}\approx -1$
(superconducting cosmic strings, for which ${\cal L}\ _{em}\approx -B^2/2)$,
or $\zeta _{CDM}\ll 1$ (neutrinos). BBN predicts an approximate value for
the baryon density of $\Omega _B\approx 0.03$ with a Hubble parameter of $%
h_0\approx 0.6$ , implying $\Omega _{CDM}\approx 0.3$. Thus depending on the
nature of the dark matter $\zeta _m$ can be virtually anything between $-1$
and $+1$. The uncertainties in the underlying quark physics and especially
the constituents of the dark matter make it difficult to impose more certain
bounds on $\zeta _m$.

We should not confuse this theory with other similar variations.
Bekenstein's theory \cite{bek2} does not take into account the stress energy
tensor of the dielectric field in Einstein's equations. Dilaton theories
predict a global coupling between the scalar and all other matter fields. As
a result they predict variations in other constants of nature, and also a
different dynamics to all the matter coupled to electromagnetism. This model
may be seen as a more conservative alternative to varying-speed-of-light
scenarios \cite{moffat93,am,ba,bm,barmag98,sn}. An interesting application
of our approach has also recently been made to braneworld cosmology by Youm 
\cite{youm}. Assuming a homogeneous and isotropic Friedmann metric with
expansion scale factor $a(t)$ and curvature parameter $k$ in eqn. (\ref{ein}%
), we obtain the field equations ($c\equiv 1$) 
\begin{eqnarray}
{\left( \frac{\dot{a}}{a}\right) }^{2} &=&\frac{8\pi G}{3}{\left( \rho
_{m}\left( 1+\mid \zeta _{m}\mid \exp {[-2\psi ]}\right) +\rho _{r}\exp {%
[-2\psi ]}+\frac{\omega }{2}\dot{\psi}^{2}\right) }  \nonumber \\
&&\ \ -\frac{k}{a^{2}}+\frac{\Lambda }{3},  \label{fried1}
\end{eqnarray}
where $\Lambda $ is the cosmological constant. For the scalar field we have
the propagation equation, 
\begin{equation}
\ddot{\psi}+3H\dot{\psi}=-\frac{2}{\omega }\exp {[-2\psi ]}\zeta _{m}\rho
_{m},  \label{psidot}
\end{equation}
where $H\equiv \dot{a}/a$ is the Hubble expansion rate$.$ Note that the sign
of the evolution of $\psi $ is dependent on the sign of $\zeta _{m}$. Since
the observational data is consistent with a {\em smaller} value of $\alpha $
in the past, we will in this paper confine our study to {\em negative}
values of $\zeta _{m}$, in line with our recent discussion in Ref.\cite{sbm}%
. The conservation equations for the non-interacting radiation and matter
densities are 
\begin{eqnarray}
\dot{\rho _{m}}+3H\rho _{m} &=&0 \\
\dot{\rho _{r}}+4H\rho _{r} &=&2\dot{\psi}\rho _{r}.  \label{conservation}
\end{eqnarray}
and so $\rho _{m}\propto a^{-3}$and $\rho _{r}$ $e^{-2\psi }\propto a^{-4},$
respectively. If additional non-interacting perfect fluids satisfying
equation of state $p=(\gamma -1)\rho $ are added to the universe then they
contribute density terms $\rho \propto a^{-3\gamma }$ to the RHS of eqn. (%
\ref{fried1}) as usual.

\section{Inhomogeneous solutions with varying $\protect\alpha $}

The Friedmann models with varying $\alpha $ have shown that when $\zeta
_{m}<0$ the homogeneous motion of the $\psi $ does not in general create
significant metric perturbations at late times and we can safely assume that
the expansion scale factor is that of the usual Friedmann universe for the
appropriate fluid. The behaviour of $\psi $ then follows from a solution of
the $\psi $ conservation equation in which the expansion scale factor is
taken to be that of the Friedmann universe for matter with the same equation
of state in general relativity ($\psi =\zeta =0$). Our earlier analyses
found that $\psi \ $ is approximately constant during the radiation era, and 
$\alpha $ increases as $2N\ln (t)$ during the dust dominated era when
spatial curvature is negligible, and tends to a constant in any subsequent
era dominated by negative spatial curvature or a positive cosmological
constant \cite{bsm}. When $\zeta <0$ we can use the same test-motion
approach to investigate inhomogeneous variations in $\psi $ and $\alpha $ as
the universe expands.

We assume that the expansion scale factor is that of the Friedmann model 
\begin{equation}
a=t^{n}  \label{a}
\end{equation}
and solve the wave equation in one of its appropriate forms: 
\begin{eqnarray}
\Box \psi  &=&-\frac{2\zeta }{\omega }\rho _{m}\exp [-2\psi ]  \label{b} \\
\ddot{\psi}+\frac{3\dot{a}}{a}\dot{\psi}-\frac{1}{a^{2}}\nabla ^{2}\psi  &=&-%
\frac{2\zeta }{\omega }\rho _{m}\exp [-2\psi ] \\
\frac{d}{dt}\left( \dot{\psi}a^{3}\right) -a\nabla ^{2}\psi  &=&N\exp
[-2\psi ]  \label{k} \\
&&\ \   \nonumber
\end{eqnarray}
where $N$ is a constant, defined by 
\[
N\equiv -\frac{2\zeta _{m}}{\omega }\rho _{m}a^{3}>0.
\]

We can see in a general way that the \ effects of small inhomogeneities in
the density of electromagnetically-coupled matter, will create a spatial
variation in $N(\vec{x})$\ and this will create small spatial variations in $%
\alpha \sim 2N(\vec{x})\ln t$\ during the dust era. It is also possible for
inhomogeneity in $N$ to be created by variations in the value of $\zeta $
for the form of matter that dominates on any particular scale. Our
assumption of $\zeta _{m}<0$ applies only to the dominant dark matter. On
small scales the luminous matter might dominate and there will be a
variation in the effective value (and even the sign) of $\zeta $ but we will
not explore these possibilities further here. 

We seek a general solution of (\ref{k}) of the form 
\begin{equation}
\psi =\psi _{h}+\delta (\vec{x},t)  \label{an}
\end{equation}
where $\psi _{h}(t)$ is the solution to the space-independent problem ($%
\nabla \psi \equiv 0$), so by definition $\psi _{h}(t)$ is an exact solution
of 
\[
\frac{d}{dt}\left( \dot{\psi}_{h}a^{3}\right) =N\exp [-2\psi _{h}] 
\]

We note immediately an important general property of this equation, that
applies to all Friedmann universes with varying $\alpha :$

{\em No-oscillation theorem:} {\em \ In the BSBM theory}$,${\em \ }$\alpha $%
{\em \ cannot display oscillatory behaviour in time in a Friedmann universe
of any curvature.}

The proof is simple: When $N$ is positive (negative) the right-hand side of
equation (\ref{psidot}) is positive (negative), $\psi _{h\text{ \ }}$cannot
have an expansion maximum (minimum) since $\dot{\psi}_{h}=0$\ and $\ \ddot{%
\psi}_{h}<0$\ ($>0$) there. Therefore $\psi _{h\text{ \ }}$cannot oscillate
in time and so neither can $\alpha =\exp [2\psi ]$. 

We see that in the case of interest, when $N>0$, $\psi $\ can have a minimum
but thereafter it must always increase irrespective of the behaviour of the
expansion scale factor. However, if the equation is linearised in $\psi _{h}$%
\ this is no longer true if attention is not confined to the small $\psi $\
regime where $\exp [-2\psi _{h}]\approx 1-2\psi _{h}\ >0$\ and spurious
oscillations of $\psi $\ (and $\alpha $) in time can appear to arise at late
times if $\psi $\ grows. It is of particular interest that this proof that $%
\psi _{h}$ cannot have a maximum applies to recollapsing universes ($k=+1$)
as well as to ever-expanding universes ($k\leq 0$). It also means that
oscillations of $\alpha $\ with redshift should not be observed in Friedmann
universe. This might prove an interesting prediction for future observations
to test.

Substituting (\ref{an}) into (\ref{k}) we get 
\[
\frac{d}{dt}\left( \dot{\delta}a^{3}\right) -a\nabla ^{2}\delta =N\exp
[-2\psi _{h}]\{\exp [-2\delta ]-1\}
\]
So for small $\delta $ 
\[
\frac{d}{dt}\left( \dot{\delta}a^{3}\right) -a\nabla ^{2}\delta =-2N\delta
\exp [-2\psi _{h}]+O(\delta ^{2})
\]
Now look for separable solutions 
\[
\delta =T(t)D(\vec{x})
\]
and we have 
\begin{equation}
\frac{\ddot{T}}{T}a^{2}+3a\dot{a}\frac{\dot{T}}{T}+\frac{2N}{a\ }\exp
[-2\psi _{h}]=-\mu ^{2}=\frac{\nabla ^{2}D}{D}  \label{T}
\end{equation}
where $\mu ^{2}$ is a separation constant with a sign chosen to ensure
non-growing, oscillatory, inhomogeneity in $D(\vec{x})$ at spatial infinity$%
\ .$ In this equation we can always neglect $\ 2Na^{-1}\exp [-2\psi _{h}]$
with respect to $\mu ^{2}$ as $t\rightarrow \infty $ because $\psi _{h}$
never falls with time (in the dust era $\psi _{h}$ grows as $\frac{1}{2}\ln
[2N\ln (t)]$ as $\ t\rightarrow \infty $, for example, \cite{bsm}). This is
an important feature of the variation of $\psi $, and $\alpha $, in BSBM
varying-$\alpha $ theories when $\zeta <0$. It ensures that the kinetic term
and the $\zeta _{m}\exp [-2\psi ]$ terms can be neglected in the Friedmann
equation asymptotically and the expansion scale factor can self-consistently
be assumed to be of the same form as when $\alpha $ does not vary (this is 
{\it not} true if $\zeta >0$). Note that in the Friedmann case ($\delta =0)$
we \ can evaluate the corrections to the test-motion approximation by
calculating the leading order corrections to the Friedmann equation if we
use the solution for $\psi $ found from the solution of the wave equation.
These are largest for the dust universe but even there we find that the next
order correction to the first-order assumption that $a(t)=t^{2/3}$ is $%
a(t)=t^{2/3}(\ln t)^{\ \left| \zeta \right| /3}$\ with $\left| \zeta \right|
/3\sim 0.3-0.03$ and so are small$.$

Hence, in this approximation we have 
\begin{equation}
\ddot{T}+\frac{3\dot{a}}{a}\dot{T}+\frac{\mu ^{2}T}{a^{2}}=0  \label{an1}
\end{equation}
and 
\[
\nabla ^{2}D=-\mu ^{2}D
\]
so we have the standard separable spherical oscillator solution in spherical
polar coordinates: 
\[
D(r,\theta ,\varphi )=\sum\limits_{\ell =0}^{\infty }c_{\mu ,\ell }Y_{\ell
}(\theta ,\varphi )r^{-1/2}Z_{\ell +\frac{1}{2}}(\mu r)
\]
where $Z$ is a cylindrical function and $Y$ the spherical harmonic function.
If we specialise to spatially-flat cosmologies with perfect fluid equations
of state for pressure $p$ and density $\rho $ of the form 
\[
p=(\gamma -1)\rho ,
\]
then the expansion scale factor will have power law form 
\begin{equation}
a=t^{n}  \label{gr}
\end{equation}
with $n=2/3\gamma .$ In these cases we have 
\begin{equation}
t\ddot{T}+3n\dot{T}+\mu ^{2}Tt^{1-2n}=0  \label{t}
\end{equation}
We are interested in inhomogeneous solutions which introduce new behaviour
as a result of including inhomogeneity. Therefore we impose a boundary
condition that $T=0$\ for $\mu =0$\ since when $D=0$\ the time-dependent
solution is already included in $\psi _{h}(t).$ Thus, for $n<1:$ 
\begin{eqnarray}
T(t) &=&t^{(1-3n)/2}C_{1}Z_{\nu }\left( \frac{\mu }{{1-n}}t^{1-n}\right) \ ,
\label{sol1} \\
\nu  &\equiv &\frac{\left| 1-3n\right| }{2(1-n)}  \label{sol2}
\end{eqnarray}
where $Z(..)$\ is a cylindrical function, while for the curvature-dominated
expansion with $n=1:$ 
\begin{eqnarray}
T &\propto &t^{q}  \label{sol3} \\
q &=&-1\pm \sqrt{1-\mu ^{2}}  \label{sol4}
\end{eqnarray}
and we choose the + solution to satisfy the boundary condition.{\em \ }The
late-time behaviour is easily determined as $t\rightarrow \infty $: 
\begin{eqnarray}
T(t) &\propto &t^{-n}\times oscillations;\hspace{1in}n\neq 1.  \label{sol5}
\\
T(t) &\propto &t^{-1+\sqrt{1-\mu ^{2}}}\ ;\hspace{1in}n=1  \label{sol6} \\
&&\ \   \nonumber
\end{eqnarray}
and decays, $T\propto a^{-1},$ as $t\rightarrow \infty .$ However, as we
have already pointed out the oscillatory behaviour is an artefact of the
linearisation process and the Bessel-like oscillation are not reached by the
solution for $\psi .$

In the radiation era we can find an solution of eqn. (\ref{T}) for $T(t)$
without neglecting the term $2Na^{-1}\ \exp [-2\psi _{h}]$ since the
radiation universe has the simple exact solution: 
\begin{equation}
\psi _{h}=\frac{1}{2}\log (8N)+\frac{1}{4}\log (t)  \label{ex}
\end{equation}
Substituting (\ref{ex}) in eqn. (\ref{T}) we find 
\[
T(t)=\frac{1}{t^{1/4}}\{AJ_{m}(2\mu t^{1/2})+BJ_{-m}(2\mu t^{1/2})\}
\]
where 
\[
m=\frac{i\sqrt{3}}{2}
\]
and we see explicitly that there is agreement with the asymptote (\ref{sol5}%
) of the approximated equation when $n=1/2$. The boundary condition for
transition to the homogeneous problems requires that we put $B=0$\ and again
the late-time oscillations are recognised as arising purely from the
linearisation process. Similar exact solutions can be found for all
universes with $1/3<n<2/3.$

The cosmological constant case of $\gamma =0$ is distinct, with 
\[
a=\exp [H_{0}t] 
\]
which gives 
\[
0=\ddot{T}+3H_{0}\dot{T}+\mu ^{2}T\ \exp [-2H_{0}t]\approx \ddot{T}+3H_{0}%
\dot{T} 
\]
as $t\rightarrow \infty ,$ so 
\[
T\rightarrow T_{\infty }-\frac{1}{3H_{0}}\exp [-3H_{0}(t+t_{0})]\rightarrow
T_{\infty } 
\]
This behaviour is in accord with the expectations of a cosmic no hair
theorem. It means that if a period of inflation occurs in the very early
universe then large scale inhomogeneity will appear increasingly negligible
with time within the event horizon of a geodesically moving observer. In the
late stages of a universe like our own, which displays evidence of being
accelerated by the presence of a positive cosmological constant, \cite{super}%
, it ensures that time variations in $\alpha $ will not grow. This is to be
expected since the inhomogeneities in density are also prevented from
growing by the effects of the cosmological constant.

\section{The case of $\protect\zeta >0$}

When the dark matter is dominated by electric field energy, we have $\zeta \
>0$, and the behaviour of eq. (\ref{psidot}) is very different to that
obtained when $\zeta <0$. Most crucially, the test-motion approximation used
above to analyse the behaviour or (\ref{psidot}) does not apply, even for
the purely time-dependent $\psi $ evolution in a Friedmann universe. The
solutions obtained for $\psi $ by assuming the scale factor evolution $%
a=t^{n}$ of general relativity (with constant $\alpha $) lead to solutions
for $\ \psi $ (and $\alpha )$ which do not increase with time. For example,
we have $\alpha \propto t^{-1}$ in the curvature era and $\alpha \propto \ln
(t_{0}/t)$ in the dust era. These contribute kinetic ($\dot{\psi}^{2}$) and
magnetic contributions ($\zeta \exp [-2\psi ]$) terms which dominate the
underlying Friedmann equation, (\ref{fried1}), at large times and the
expansion of the universe is not well approximated by that obtained in
general relativistic cosmologies with the same equation of state and
constant $\alpha $ except over finite non-asymptotic intervals of time. This
leads to problems accommodating observational constraints, notably the
results of studies of the structure of the microwave background at last
scattering {\cite{avelino,bat} }and big bang nucleosynthesis \cite{bbn} in
the radiation era because the value of $\alpha $ then is significantly
different from today, unlike in the cases of $\zeta <0$. Cosmologies with $%
\zeta >0$ have been discussed in ref. \cite{olive} in a theory that is
similar in structure to the BSBM theory discussed here. We will discuss the $%
\zeta >0$ version of the theory in more detail elsewhere. It is less well
behaved and does not seem to provide the smooth and simple perturbation of
the standard cosmology with constant $\alpha $ as seen in the negative $%
\zeta $ case.


\section{Discussion}

We have shown that the time-dependent solutions to the Friedmann model are
stable against the effects of inhomogeneous motions of the $\psi $ field. In
the case of inhomogeneous variation the cosmological solutions in universes
with scale factor $a(t)=t^{n}$, to leading order take the form: 
\[
\psi (\vec{x},t)=\psi _{h}(t)+C_{1}t^{(1-3n)/2}\left[ J_{\nu }\left( \frac{%
\mu }{{1-n}}t^{1-n}\right) \ \right] \sum\limits_{\ell =0}^{\infty }c_{\mu
,\ell }Y_{\ell }(\theta ,\varphi )r^{-1/2}Z_{\ell +\frac{1}{2}}(\mu r) 
\]
when $n\neq 1$, and

\[
\psi (\vec{x},t)=\psi _{h}(t)+At^{-1+\sqrt{1-\mu ^{2}}}\ \sum\limits_{\ell
=0}^{\infty }c_{\mu ,\ell }Y_{\ell }(\theta ,\varphi )r^{-1/2}Z_{\ell +\frac{%
1}{2}}(\mu r) 
\]
when $n=1$, while for the case of $a=\exp [H_{0}t]:$ 
\[
\psi (\vec{x},t)=\psi _{h}(t)+O(\exp [-3H_{0}t\ ]).\hspace{1in} 
\]
Thus in all cases we have 
\[
\psi (\vec{x},t)\rightarrow \psi _{h}(t) 
\]
as $t\rightarrow \infty $ and at late times spatial variations in the fine
structure constant decay as 
\begin{eqnarray*}
&& \\
\alpha &=&\exp [2\psi _{h}]\{1+2t^{-n}\sum\limits_{\ell =0}^{\infty }c_{\mu
,\ell }Y_{\ell }(\theta ,\varphi )r^{-1/2}Z_{\ell +\frac{1}{2}}(\mu r)\ +..\}
\end{eqnarray*}
for $n\neq 1.$ Hence, denoting $\alpha _{h}\equiv \exp [2\psi _{h}],$ the
spatial variation in $\alpha $ decays in time in the $n\neq 1$ universes as 
\[
\frac{\delta \alpha }{\alpha }\equiv \frac{\alpha -\alpha _{h}}{\alpha _{h}}%
\approx 2t^{-n}\sum\limits_{\ell =0}^{\infty }c_{\mu ,\ell }Y_{\ell }(\theta
,\varphi )r^{-1/2}Z_{\ell +\frac{1}{2}}(\mu r)\ 
\]
Analogous expressions can be written down mutatis mutandis for $\delta
\alpha /\alpha $ in the $n=0,1$ cases.

It is important to compare the evolution of the fine structure constant $%
\alpha (t)$ in the BSBM theory in the homogeneous case with that for the
situation admitting inhomogeneous motions of the fine structure 'constant', $%
\alpha (t,\vec{x}),$ here. To leading order, the overall pattern of time
evolution studied in refs. \cite{sbm,bsm} is unaffected by the presence of
small inhomogeneities. However, small spatial variations of an oscillatory
character are expected to exist in the value of the fine structure constant
over astronomical scales, reflecting the non-linear self-interaction of the $%
\alpha (\psi )$ field which carries the variations in $\alpha .$ The spatial
variation amplitudes, $\delta \alpha /\alpha $, are found to decay with time
as the universe expands and will not be as significant as the overall
variation in time of the mean value of $\alpha (t)\propto \ln (t)\ $ during
the dust-dominated phase of a spatially-flat universe. Inhomogeneous test
motions of the $\psi $ field will have been decaying in amplitude throughout
the period when the universe was dominated by dust if $\zeta <0.$ Therefore
we would not expect any significant inhomogeneities to survive at the
astronomically interesting epoch $z\sim 1-4$ where the value of the fine
structure constant can be probed spectroscopically with high precision.
However, our discussion has not considered three situations where more
significant spatial variations might arise. The first is the situation
within gravitationally bound matter inhomogeneities of large scale which
separate out from the expansion of the Universe and collapse to form
superclusters and clusters of galaxies. These behave in a manner similar to
that expected of separate closed universes until deviations from spherical
symmetry become significant. Our analysis is not applicable here because the
dynamics of the bound inhomogeneities will differ significantly after they
separate off from the background expansion. The second situation of interest
is that in which perturbations of the Friedmann metric are included in the
problem and allowed to couple to spatial variations in $\psi ,$ or $\alpha .$
This coupling will lead to small temperature fluctuations in the microwave
background radiation. Finally, the variation in the value and sign of $\zeta 
$ with scale for the dominant form of matter could introduce a distinctive
inhomogeneity. These problems will be discussed elsewhere.

Acknowledgements We would like to thank David Mota for discussions.


\begin{references}
\bibitem{sbm}  H. Sandvik, J.D. Barrow, and J. Magueijo, Phys. Rev. Lett.
88, 031302 (2002).

\bibitem{bsm}  J.D. Barrow, H. Sandvik, and J. Magueijo, Phys. Rev. D 65,
063504 (2002) and Phys. Rev. D 65, 123501 (2002)

\bibitem{bek2}  J.D. Bekenstein, Phys. Rev. D 25, 1527 (1982).

\bibitem{webb}  J.K. Webb, V.V. Flambaum, C.W. Churchill, M.J. Drinkwater \&
J.D. Barrow, Phys. Rev. Lett. 82, 884 (1999).

\bibitem{webb2}  J.K. Webb et al. Phys. Rev. Lett. 87, 091301 (2001),
astro-ph/0012539

\bibitem{murphy}  M.T. Murphy et. al., Mon. Not. Roy. Astron. Soc. 327, 1208
(2001).

\bibitem{sh}  A. Shlyakhter, Nature 264, 340 (1976); ATOMKI Report A/1,
(1983).

\bibitem{fujii}  Y. Fujii et al, Nucl. Phys. B, 573, 377 (2000).

\bibitem{avelino}  P.P. Avelino et. al., Phys. Rev. D 62, 123508, (2000) and
astro-ph/0102144 (2001).

\bibitem{bat}  R. Battye, R. Crittenden, J. Weller, Phys. Rev. D 63, 043505,
(2001).

\bibitem{bbn}  J.D. Barrow, Phys. Rev. D, 35, 1805 (1987).

\bibitem{ivan}  A.V. Ivanchik, E. Rodriguez, P. Petitjean, and D.A.
Varshalovich, astro-ph/0112323.

\bibitem{bt}  J.D. Barrow and C. O'Toole, Mon. Not. R. astron. Soc. 22, 585
(2001).

\bibitem{am}  A. Albrecht and J. Magueijo, Phys. Rev. D 59, 043516 (1999).

\bibitem{covvsl}  J. Magueijo, Phys. Rev. D 62, 103521, (2000).

\bibitem{moffatal}  J. Moffat, astro-ph/0109350.

\bibitem{land}  S. Landau, P. Sisterna \& H. Vucetich, Phys. Rev. D63,
081303(R) (2001)

\bibitem{zal}  G. Dvali and M. Zaldarriaga, hep-ph/0108217.

\bibitem{olive}  K. Olive and M. Pospelov, hep-ph/0110377.

\bibitem{livio}  M. Livio and M. Stiavelli, Ap.J. Lett. 507, L13 (1998);
Cowie L. and Songaila A., ApJ, 453, 596 (1995); Varshalovich D. A. and
Potekhin A. Y., Space Sci. Rev., 74, 259 (1995).

\bibitem{youm}  D. Youm, hep-th/0108237.

\bibitem{moffat93}  J. Moffat, Int. J. Mod. Phys. D 2, 351 (1993).

\bibitem{ba}  J.D. Barrow, Phys. Rev. D 59, 043515 (1999).

\bibitem{bm}  J.D. Barrow and J. Magueijo. Phys. Lett. B 447, 246 (1999).

\bibitem{barmag98}  J.D. Barrow and J. Magueijo, Phys. Lett. B{\bf \ } 443,
104 (1998).

\bibitem{sn}  J. D. Barrow and J. Magueijo, Ap. J. Lett. 532, L87 (2000).

\bibitem{super}  S. Perlmutter et al., Ap. J. 517, 565 (1999); S. Perlmutter
et al, Ap. J. 483, 565 (1997); S. Perlmutter et al, Nature 391 51 (1998);
P.M. Garnavich, et al, Ap.J. Letters 493, L53 (1998); Schmidt, B.P., Ap.J.
507,46 (1998); Riess, A.G. et al 1998 AJ 116,1009.
\end{references}
\end{document}